\documentclass[pdflatex,sn-mathphys-num, referee]{sn-jnl}

\usepackage{graphicx}%
\usepackage{multirow}%
\usepackage{amsmath,amssymb,amsfonts}%
\usepackage{amsthm}%
\usepackage{mathrsfs}%
\usepackage[title]{appendix}%
\usepackage{xcolor}%
\usepackage{textcomp}%
\usepackage{manyfoot}%
\usepackage{booktabs}%
\usepackage{algorithm}%
\usepackage{algorithmicx}%
\usepackage{algpseudocode}%
\usepackage{listings}%
\usepackage{float}

\theoremstyle{thmstyleone}%
\theoremstyle{thmstyletwo}%

\theoremstyle{thmstylethree}%

\begin{document}

\title[Avoiding Overfitting in Variable-Order Markov Models: a Cross-Validation Approach]{Avoiding Overfitting in Variable-Order Markov Models: a Cross-Validation Approach}


\author*[1]{\fnm{Valeria} \sur{Secchini}}\email{valeria.secchini.r@gmail.com}

\author[2,3]{\fnm{Javier} \sur{Garcia-Bernardo}}\email{j.garciabernardo@uu.nl}

\author[1]{\fnm{Petr} \sur{Janský}}\email{petr.jansky@fsv.cuni.cz}

\affil*[1]{\orgdiv{CORPTAX, Institute of Economic Studies, Faculty of Social Sciences}, \orgname{Charles University}, \orgaddress{\city{Prague}, \country{Czech Republic}}}

\affil[2]{\orgdiv{Department of Methodology and Statistics}, \orgname{Utrecht University}, \orgaddress{\city{Utrecht},\country{The Netherlands}}}

\affil[3]{\orgdiv{Centre for Complex Systems Studies}, \orgname{Utrecht University}, \orgaddress{\city{Utrecht},\country{The Netherlands}}}


\abstract{Higher-order Markov chain models are widely used to represent agent transitions in dynamic systems, such as passengers in transport networks. They capture transitions in complex systems by considering not only the current state but also the path of previously visited states. For example, the likelihood of train passengers traveling from Paris (current state) to Rome could increase significantly if their journey originated in Italy (prior state). Although this approach provides a more faithful representation of the system than first-order models, we find that commonly used methods-relying on Kullback–Leibler divergence-frequently overfit the data, mistaking fluctuations for higher-order dependencies and undermining forecasts and resource allocation. Here, we introduce DIVOP (Detection of Informative Variable-Order Paths), an algorithm that employs cross-validation to robustly distinguish meaningful higher-order dependencies from noise. In both synthetic and real-world datasets, DIVOP outperforms two state-of-the-art algorithms by achieving higher precision, recall, and sparser representations of the underlying dynamics. When applied to global corporate ownership data, DIVOP reveals that tax havens appear in 82\% of all significant higher-order dependencies, underscoring their outsized influence in corporate networks. By mitigating overfitting, DIVOP enables more reliable multi-step predictions and decision-making, paving the way toward deeper insights into the hidden structures that drive modern interconnected systems.}

\vspace{1cm}


\maketitle

\section{Introduction}\label{sec1}

Dynamical systems, such as ownership relationships among firms \cite{garcia2017}, passengers travelling through airports \cite{hertzberg2018behaviors}
or protein-protein interactions \cite{Vazquez2003}, are commonly modeled as first-order Markov models, where nodes represent states (e.g. airports or countries), and edges represent transitions (e.g. flights or cross-border ownership).
We refer to this data as \textit{sequences} data---a series of states representing a transition of agents between those states.
To illustrate, we consider passengers flying through countries. Passengers leaving from the United Kingdom could have a certain probability to go to Italy. 
An option to model dynamical systems is using a Markov Chain Model (MCM), where connections between \textit{states} (in our example, the country)  in the model represent transition probabilities between the states (in our example, passenger flows).
In a first order MCM, it is completely irrelevant if the passengers arrived in the United Kingdom from the United States or from France. 
However, this is not realistic. Passengers that arrived in the United Kingdom from the United States may be in a connecting flight to Italy. But given the availability of direct trains and flights from France to Italy, passengers that arrived in the United Kingdom from France are very unlikely to fly to Italy. Since the probability of flying from the United Kingdom to Italy depends on the previous departure, it is said that the system has memory and that first-order Markov models do not accurately reflect the dynamics of the system \cite{salnikov2016using,Xu2016,Rosvall2014,ganmor2011sparse,Vroylandt2022Likelihood,ayaz2021non}.

To include memory and accurately reconstruct the dynamics of systems, a growing body of literature focuses on higher-order Markov models\cite{Lambiotte2019}. Relaxing the first-order assumption can profoundly alter network properties and dynamics \cite{BENSON2019,Rosvall2014,lambiotte2015effect, Scholtes2014}.
A first-order MCM estimates the transition probabilities of going from a single state to another (UK, USA, Italy and France in our example).
A second-order MCM, instead, contains every combination of pair states as nodes of the model. In our example, the nodes of the model would be (France,UK), (USA,UK), (UK,Italy), etc. 
This allows us to assign different probabilities to flying from the UK to Italy if the passengers arrive in the UK from France (France,UK) or from the United States (USA,UK). In the case of a third-order MCM, 
the nodes would consist, instead, of combinations of three states (e.g. (USA,UK,Italy)) and so on.

Several methods have been developed in the literature to distinguish which order fits the data better, such as models based on the
likelihood ratio (LR) \cite{Anderson1957,menendez2001,Scholtes2017,Tong1975,Akaike1974}, Bayesian approaches \cite{Luka2022,Strelioff2007,Singer2015, Peixoto2017,katz1981some}, or methods based on mutual information \cite{Papapetrou2013,Papapetrou2016} or chi-square statistics \cite{hiscott1981chi,baigorri2014markov,pethel2014exact}. An interesting network approach to represent the whole higher order MC  can be found in\\
\cite{Scholtes2017}, where all the nodes (combination of states) of each order are\\
contained in a different layer of a multilayer network \cite{Mikko2014}.

However, the limitation of these higher-order Markov chains is that the number of nodes grows exponentially with the estimated order $k$ of the MCM ---as $O(N^k)$ where $N$ is the number of states.
Especially in large datasets, this quickly becomes unfeasible.
Modeling a process of 1,000 states using third-order interactions results in a final network of 1 billion nodes.

Methods have been developed to reduce the model dimension, for example, relying on Sparse Markov Models \cite{jaaskinen2014sparse} or on the construction of variable order Markov models (VOM). The most used VOMs are based on the ``pruning'' the higher order nodes that are not changing the dynamics of the Markov model in comparison with the previous (lower order) step. 
In our example, this would involve calculating the probability distribution of the route $(USA,UK) \rightarrow C$, where $C$ is all possible countries reachable from the UK, and comparing it with the probability distribution of the route $UK \rightarrow C$ in the first-order frame.  
If these two transition probabilities are similar, we can then prune the node (USA,UK), given that it does not add additional information regarding the dynamics. 
The ``similarity'' between the two probabilities is usually based on the Kullback-Leibler (KL) divergence, which measures the divergence between the lower and higher order transition probability distributions ($\Delta$). The method was introduced in \cite{rissanen1983universal} by Rissanen and later used and further developed in
various works \cite{Shmilovici2007,Buhlmann1999, Xu2016, Saebi2020, Ben2005, Schreiber2000,orlov2002construction, weinberger1995universal,ben2003context,vert2001adaptive}.
These pruned Markov models have usually been depicted in the so called \textit{context trees} \cite{rissanen1983universal}. However, another convenient and novel representation was given in \cite{Xu2016}, where they portray the higher and lower order nodes in a unique network.
Another potential method to creating variable order Markov models (VOM) is the HYPA algorithm, developed recently by \cite{Larock2020}. 
The HYPA algorithm,  instead of the KL divergence, finds ``anomalous'' higher-order nodes when comparing them with a null model, built from the previous order.
Other methods to estimate VOM models are based on a Bayesian approach \cite{pmlr-v9-dimitrakakis10a, begleiter2004prediction}. 

In this paper, we show that traditional KL divergence models tend to overfit the data, and we present our model that substantially reduces it. Indeed, in the previous pruning models, nodes that should be pruned are left in the model because fluctuations in the data are wrongly detected as divergence between the higher- and lower-order distributions. To solve this problem, we introduce and incorporate a quantity to measure the noise inside the distributions, $V$, through cross-validation, thus taking into account that fluctuations can alter the estimation of the $\Delta$.

Overfitting includes fluctuations in the dynamics reconstruction, which hinders the forecasting of transitions and the allocation of resources in the most efficient way. For example, it could include tens of thousands of higher-order paths in the dynamics of flights in airport, which would make the organization of flight transfers unfeasible.

We build on methods based on the KL divergence, in particular the BuildHON model \cite{Xu2016}, adding a further quantity in the model, a measure of the noise of higher orders paths'.
This massively reduces overfitting and yields sparser variable-order MCM---i.e., random noise in the data is less likely to be detected as a significant KL divergence. We called our algorithm DIVOP (Detection of Informative Variable Order Paths). 
We test our model and compare it with two recently developed algorithms, BuildHON+ \cite{Saebi2020} and HYPA \cite{Larock2020}, on three different synthetic datasets. Our model increases both precision and recall, especially for orders higher than two. 
From further analysis, our novel measure of noise proves to be a robust and accurate measure of fluctuations in the datasets. 
We find that divergence ($\Delta$), variability ($V_P$) and frequency of appearance of higher-order nodes in the data interact between each other, preventing us from using a simple threshold to determine if the higher-order node is necessary. 
Instead, we create synthetic data resembling real-world datasets. This allows us to match the thresholds to the specific dataset to be studied. Finally, we apply our algorithm to a dataset containing corporate ownership sequences---in which a series of companies are connected via ownership relations---of around $18'000$ major multinational corporations. We also test the model on the dataset through a random walk test, and find that it significantly improves the forecasting of the dynamics in comparison with the first-order and at the same level of HYPA.
We show that considering higher-order dependencies highlights the role of tax havens in the organization of multinational corporations and reconstructs the dynamics in a more reliable way in comparison with the first-order. Our code is published open-source at xxxx [to add after acceptance].

The rest of the paper is organized as follows. In section~\ref{sec:methods}  we introduce our model detecting \textit{informative paths}: higher-order nodes that provide extra information in comparison with their corresponding lower-order nodes. In section~\ref{sec:results} we show the performance of DIVOP in terms of precision and recall, comparing it to the BuildHON \cite{Xu2016} and HYPA \cite{Larock2020} algorithms in both synthetic and real data. Section~\ref{sec:conclusion} concludes by summarizing our results and suggesting implications for further research.

\section{Methods}\label{sec:methods}

\subsection{DIVOP: Detecting Higher-Order Informative Paths in Sequences} \label{sec:methods:our_model}
\begin{figure}[t]
    \includegraphics[width=1\textwidth]{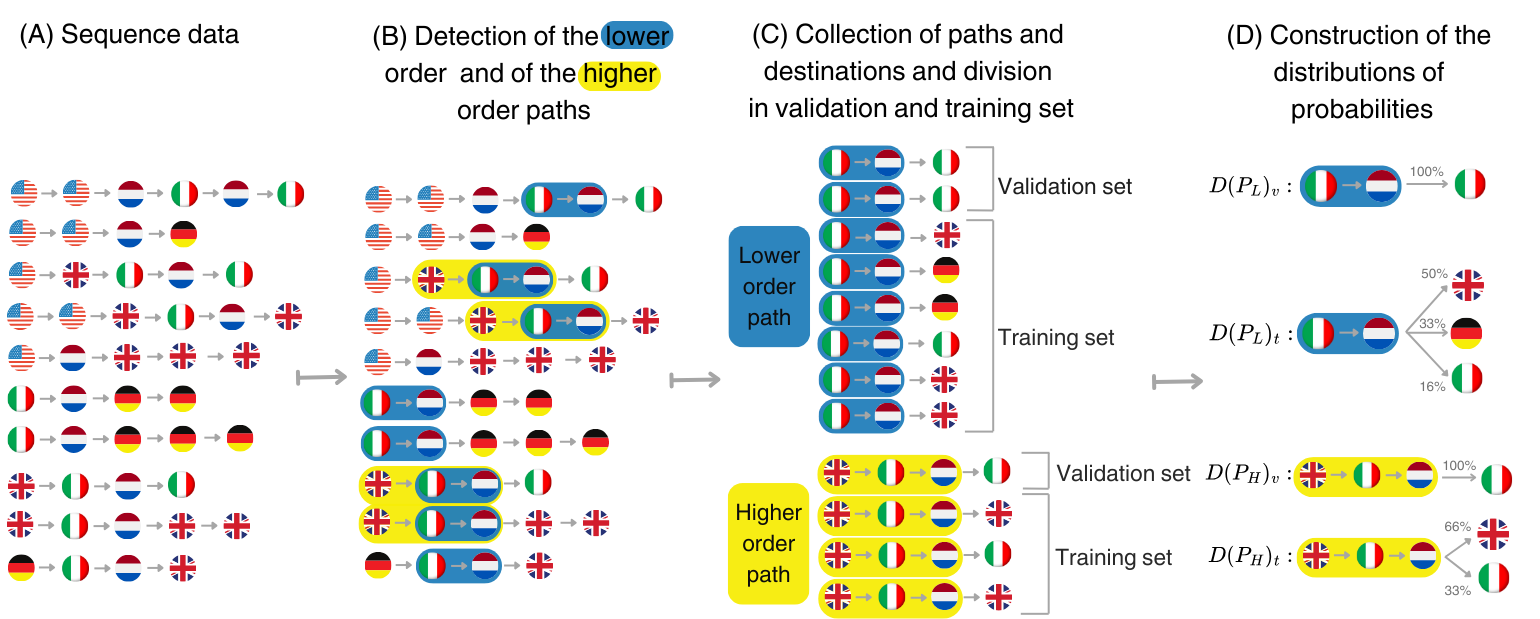}
    \caption{\textbf{Summary of our DIVOP.} (A) Sequence data representing transition between countries. (B) Example of the third-order path $(UK,IT,NL)$ and its counterpart second-order path $(IT,NL)$. The destinations of the path are the states reached after the Netherlands (NL). (C) Example division of the destinations of the path in validation and training sets using stratified cross-validation. (D) Calculation of the probability distribution of destinations. At each of the $4$-folds, a different 25\% of the data is used for validation, with the rest used for training. The same procedure is applied to every path.}
    \label{fig:division_lower}
\end{figure}

Sequence data represents  transitions between successive  states (Fig. \ref{fig:division_lower}A).
We define each sequence to be consisting of \textit{paths}, $P_i$: a series of states, $P_i = (s_1,s_2,\cdots,s_k)$, reflecting the movement of an agent through the states. The \textit{destinations} of a path are all the possible states reached by the agents after the path. For each higher-order path, $P_H$, of order $k$, $P_L$ represents its $k-1$ order counterpart (in the example of Fig. \ref{fig:division_lower}B, $P_H = (UK,IT,NL)$, $P_L = (IT,NL)$)). The destinations of $P_H$ and $P_L$ in our example are Italy, the United Kingdom, and Germany (Fig. \ref{fig:division_lower}B)

Our algorithm calculates a cross-validated divergence between the probability distribution of destinations of each $P_H$ and the probability distribution of destinations of its corresponding $P_L$. We used a stratified 4-fold cross-validation, ensuring that at each fold we sample the same proportion of destinations coming from higher and lower-order paths as in the original dataset (Fig. \ref{fig:division_lower}C). For each fold, we then construct the probability distribution of destinations in the training and validation set for both the higher-order (namely $D_{H_v}$ and $D_{H_t}$) and its corresponding lower order path (namely $D_{L_v}$ and $D_{L_t}$) (Fig. \ref{fig:division_lower}D).
If these distributions are consistently different across folds and noise is recognized as not driving this difference, we label $P_H$ as an \textit{informative path}, since it gives additional information on the dynamics of the system. The informative path will become a \textit{higher-order node} in the variable-order Markov chain. 
This allows us to construct a variable-order Markov chain containing both higher-order and first-order dependencies.

We assess if $P_H$ is an informative path using five metrics:
\begin{itemize}[leftmargin=*]
    \item \textit{$\Delta$}:
    The Jensen Shannon Divergence (JSD) between the distributions of destinations in the training dataset: $JSD(D_{H_t}|D_{L_t})$. This metric reflects the distance between the higher and lower-order distributions. A large distance signals that the higher-order interaction may be informative. However, this distance may also be driven by noise. To avoid overfitting it, we include two extra metrics.
    \item \textit{$V_{P_H}$} (``Variability of $P_H$''): 
    The distance between the distribution of destinations for the higher-order path in the validation and training set: $JSD(P_{H_t}|P_{H_v})$. A low $V_H$ indicates that the distribution of the higher-order interaction is similar in the validation and training dataset, and thus reliable. A high $V_H$ indicates that a potential difference in $\Delta$ may be driven by noise.
    \item \textit{$V_{P_L}$}: (``Variability of $P_L$''): 
    The distance between the distribution of destinations for the lower order path n the validation and training set: $JSD(P_{L_t}|P_{L_v})$. A high $V_L$ indicates that a potential difference in $\Delta$ may be driven by noise.
    \item $F_{P_H}$ The frequency of $P_H$ in the whole dataset.
    \item $F_{P_L}$ The frequency of $P_L$ in the whole dataset.
\end{itemize}

The Jensen Shannon Divergence $JSD(D_1|D_2)$ between two probability distributions $D_1$ and $D_2$ is defined as
\begin{equation*}
    JSD(D_1|D_2) = \frac{1}{2}KLD(D_1|M) + \frac{1}{2}KLD(D_2|M),
\end{equation*}

where $M=\frac{1}{2}(D_1 + D_2)$ and $KLD$ is the Kullback-Leibler divergence:

\begin{equation*}
    KLD(D_A|D_B) = \sum_i D_A(i) \log_2\biggl(\frac{D_A(i)}{D_B(i)}\biggl).
\end{equation*}

We use the \textit{JSD} instead of the \textit{KLD} directly because, due to the division into training and validation sets, the distributions $D_1$ and $D_2$ may not contain the same nodes, which could result in a division by zero. The \textit{JSD} can be seen as the divergence between each distribution and the average distribution.

The variability of the paths ($V_{P_H}$ and $V_{P_L}$) allows the algorithm to detect if the divergence between the higher- and lower-order distributions is due to random fluctuations in the data or if it is a genuine difference. It also proves to be a reliable measure of noise in the data, as can be seen in Supplementary Information  \ref{subsec:distances_analysis}

\begin{figure}[h!]
    \centering
    \includegraphics[width=0.9\textwidth]{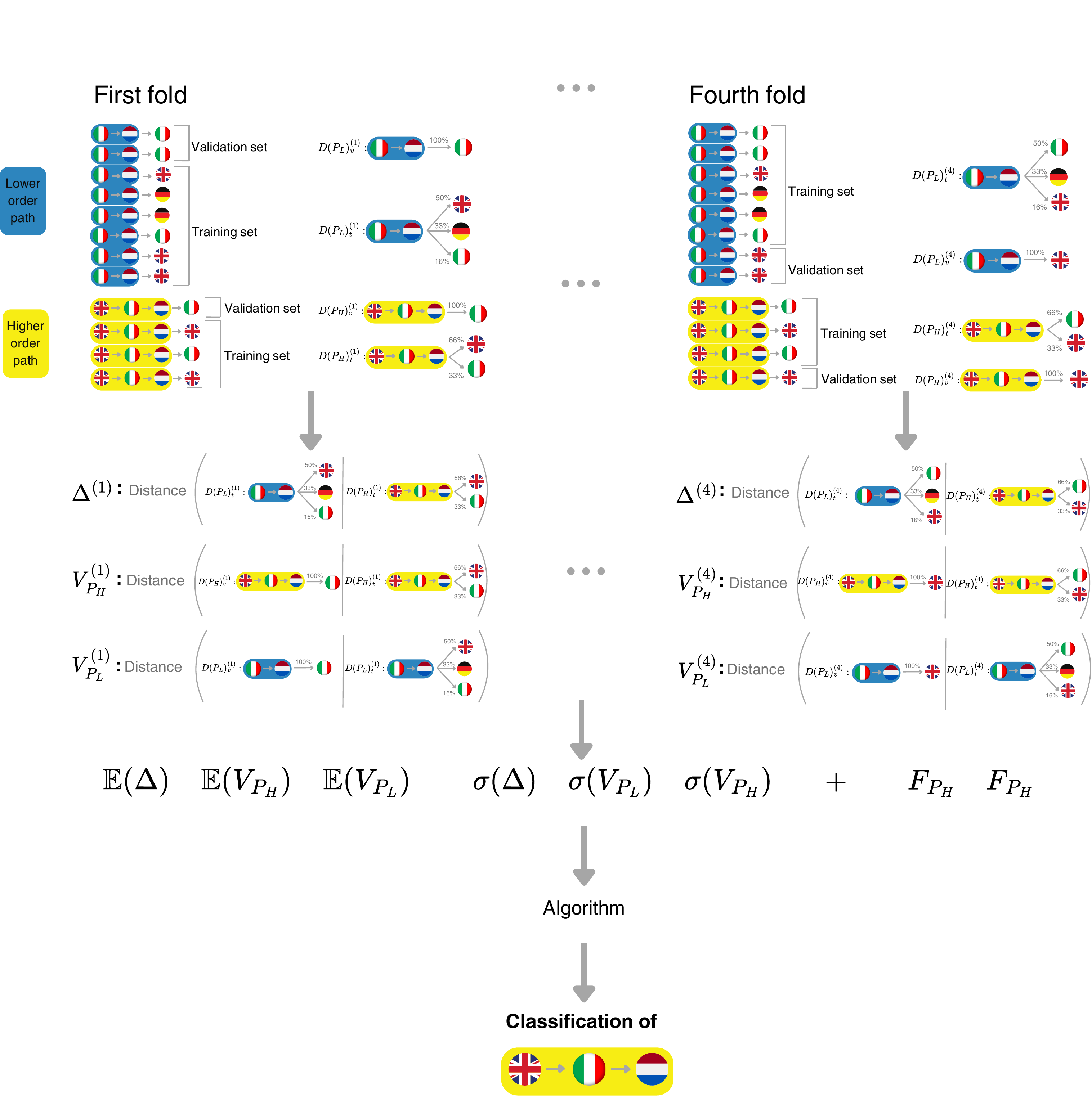}
    \caption{\textbf{Classifying paths as informative.} Taking the example of the sequence data shown in Fig. \ref{fig:division_lower}, we illustrate the classification process of the path $(UK,IT,NL)$. Once we collect the destinations in the data of the paths $(UK,IT,NL)$ and $(IT,NL)$, we divide the patterns into $4$ folds (iterations). At every fold, we take a different 25\% of the data as the validation set and the rest as the training set. At each iteration, we calculate the distributions in the validation and training set for both the lower and the higher-order. We use the distances of the distributions to calculate the variables $\Delta$, $V_{P_H}$ and $V_{P_L}$, as well as the empirical frequency of $P_H$ and $P_L$.
    We then calculate the mean and the standard deviation of the metrics on all the iterations. The values are then used as inputs of a classification algorithm to predict if the higher-order $(UK,IT,NL)$ is an informative path.
    }
    \label{fig:iterations}
\end{figure}

We finally calculate the mean and the standard deviation of the five metrics in a $4$-fold stratified cross-validation\footnote{Paths incurring less than four times are removed from the sample}. Based on those eight values, we either classify $P_H$ as an informative path, or remove it from the variable-order Markov model. We detail the classification algorithm below in Section~\ref{sec:methods:classifier}. The full process is illustrated in Figure~\ref{fig:iterations}.

\vspace{1cm}

\subsection{Classifier}\label{sec:methods:classifier}

For all algorithms, the precision, recall and F1-score of the detected higher-order nodes in synthetic data depend critically on the thresholds used. These thresholds are dataset specific and depend on the length of the sequences and the frequency of the paths.

As our classifier, we use the Histogram Boosting Gradient Classifier (HBGC) from the sckit-learn Python library\footnote{https://scikit-learn.org/stable/about.htm}. This algorithm is inspired by \textit{LightGBM}\footnote{https://lightgbm.readthedocs.io/en/stable/} and performs well for a wide range of classification tasks. We use as inputs the mean and standard deviation across folds of $\Delta$, $V_{P_H}$, and $V_{P_L}$, as well as $F_{P_H}$, and $F_{P_L}$ (Section ~\ref{sec:methods:our_model}) extracted from the synthetic data 
(Section \ref{sec:methods:syn}), for which we know which paths are informative paths.  In particular, we fit the classifier using the mean and standard deviation of $\Delta$, $V_{P_H}$, $V_{P_L}$, and the values of $F_{P_H}$ and $F_{P_H}$. The training data for the algorithm are obtained concatenating around $100$ synthetic sets of each data source. We then create 10 different samples to assess the robustness of the results and test the models on each of them individually. We tune the hyperparameters (learning rate, max depth, max iter, minimum sample leaf) to maximize the F1-score. We used the resulting fitted model to detect higher-order nodes in different synthetic data and in the real-world datasets.  

We provide code to create synthetic datasets based on a given dataset and to optimize the classifier. A key insight of the classifier is that there is not a single cut-off point that would allow us to separate true higher-order nodes from false negatives. Instead, increases in the value of different metrics increases the probability of being classified as a higher-order node gradually (see section \ref{sec:variable_analysis} in the Supplementary Information)

\subsection{Construction of Synthetic Data}\label{sec:methods:syn}

We develop a general methodology to create synthetic data that resemble the characteristics (number of nodes, number of sequences and length distribution of the sequences) of datasets but where we can control the number of synthetic higher-order nodes. 

First, we fit a first-order Markov model to the data---i.e., we calculate the normalized adjacency matrix.

To set the synthetic informative paths, we generate a path $(\cdots,i,j)$, where every state $j$ is put after state $i$ only if there is a directed edge from $i$ to $j$ in the adjacency matrix. Once $(\cdots,i,j)$ is set, we take the transition probabilities of $j$, $D(j)$, from the adjacency matrix and create $D_{(\cdots,i,j)}$, changing $D(j)$. We pick one possible destination of $D(j)$ and increase its probability to appear after $(\cdots,i,j)$ by a random number in the interval $[0.1, 1]$. We then adjust the probabilities of all the other possible destinations accordingly, so that $D_{(\cdots,i,j)}$ is normalized to sum to 1. We create a number $r$ of synthetic informative paths, for every order, until the $w^{th}$ order.

The process to then generate the synthetic sequences is as follows:
\begin{itemize}
    \item[$\boldsymbol{\cdot}$] We write one state after another, where the first state $f$ is chosen from the same distribution with which first states appear in the real sequences.
    \item[$\boldsymbol{\cdot}$]  The following node is chosen following the transition probabilities of $f$ in the adjacency matrix.
    \item[$\boldsymbol{\cdot}$] From the third node on, we check if a series of states corresponding to a synthetic informative path previously generated has appeared in the data. 
    \item[$\boldsymbol{\cdot}$] If so, the following state is chosen with a probability taken from $D_{(\cdots,i,j)}$ .
    \item[$\boldsymbol{\cdot}$] Otherwise, we will still use the adjacency matrix probability $D(j)$ (the first-order one).
\end{itemize}
In this way, the synthetic informative paths will deviate from the first-order dynamics and can be detected as such.

We create two different types of synthetic datasets per data source: the first is used to compare the performance of the three studied algorithms. We create 10 different samples to assess the robustness of the results and test the models on each of them individually. The second type of synthetic data is used to train the classifier to decide if an extracted path in the test set is a higher-order node.  In our model, this involves fitting a classifier using the eight metrics: the mean and standard deviation of $\Delta$, $V_{P_H}$, $V_{P_L}$, and the values of $F_{P_H}$ and $F_{P_H}$. In HYPA, we do this by setting the threshold (see Section~\ref{Other_models} for an explanation of the threshold) to the value that maximizes the cross-validated F1 score in the dataset. This second synthetic dataset is created concatenating $100$ synthetic sets until reaching a data size of around 4 billion observations.

\section{Results}\label{sec:results}

\subsection{Performance of DIVOP and on Synthetic Datasets}
We test the performance of DIVOP and two other recently developed models (BuildHON \cite{Xu2016} and HYPA \cite{Larock2020}, more details on the models in the Supplementary Information \ref{Other_models}) on the three synthetic datasets (\ref{sec:methods:syn}). We use all three algorithms to predict which higher-order paths are the synthetic informative paths we put in the data. Our model outperforms the F1-score of both the Saebi and HYPA algorithms in synthetic data resembling the real-world  Orbis and flights datasets (Table \ref{tab:metrics})

For both the Orbis and Flights dataset, the Saebi model has the highest recall, but at the expense of very low precision (7.5\% and 14.2\%). This implies that 85-92\% of the informative paths detected are false positives and the model is strongly overfitting the data. Consequently, the F1-score of this algorithm is the lowest by over 50 percentage points (Table \ref{tab:metrics}). The HYPA model has a similar performance to our model, with an F1 score of 64 and 87\% for the two datasets, compared to 70 and 92\% in our model (Table \ref{tab:metrics}). 

For the Xu et al. dataset, both our and BuildHON \cite{Xu2016} model perform perfectly. This is perhaps not surprising, as in this dataset the distribution of destinations is strongly changed in higher-order nodes. We lack the performance metrics for the HYPA model for this case since we have a computational constraint (due to crushing) because this dataset is very large (100,000 sequences of 100 states each). However, we expect HYPA to perform perfectly as well. The same happens with HYPA for orders higher than four in all datasets.

While BuildHON \cite{Xu2016} has a fixed (not depending on the dataset) threshold to classify the paths, for both DIVOP and HYPA \cite{Larock2020}, we use the synthetic datasets to set the weights/threshold and maximize the F1 score.

\begin{table}[htbp]
\centering
\begin{tabular}{lp{2cm}p{2cm}p{2cm}p{2cm}}
\toprule
Dataset & Metrics & BuildHON & HYPA & DIVOP \\
\midrule
\textbf{Orbis data} & Precision & $7.8 \pm 0.4\%$ & $76 \pm 2\%$ & \textbf{$89 \pm 1\%$} \\
& Recall & \textbf{$67 \pm 2\%$} & $55 \pm 3\%$ & $60 \pm 2\%$ \\
& F1 score & $13.4 \pm 0.5\%$ & $64 \pm 3\%$ & \textbf{$78 \pm 2\%$} \\
\midrule
\textbf{Flights data} & Precision & $14.2 \pm 0.4\%$ & $95.7 \pm 0.8\%$ & \textbf{$97.5 \pm 0.5\%$} \\
& Recall & \textbf{$86 \pm 1\%$} & $79 \pm 2\%$ & \textbf{$86.1 \pm 0.9\%$} \\
& F1 score & $24.4 \pm 0.7\%$ & $87 \pm 1\%$ & \textbf{$91.5 \pm 0.5\%$} \\
\midrule
\textbf{Xu et al. data} & Precision & 100\% & -\% & 100\% \\
& Recall & 100\% & -\% & 100\% \\
& F1 score & 100\% & -\% & 100\% \\
\bottomrule
\end{tabular}

\vspace{0.2cm}

\caption{Performance of the algorithms on synthetic datasets. Precision, recall, and F1-score for detecting higher-order nodes.}
\label{tab:metrics}
\end{table}

After showing that DIVOP exceeds the performance of other methods, we disaggregated the performance by the order and the frequency of the paths in the synthetic dataset.  DIVOP's performance was better than other methods across different frequencies and orders, especially when the frequency of the paths is low (less than 20 appearances) and at order 3--5. For more details see \ref{subsec:disaggregating} in the Supplementary Information.

Finally, we investigated which metrics used in the classifier ($\Delta$ and $V_P$) were able to distinguish higher order nodes. As expected, the distance is higher in the second order paths marked as informative (see Fig.\ref{fig:D_O}), while the average $\Delta$ is quite similar between positives and negatives for orders higher than two. This is why the variability is crucial to distinguish positives and negatives, given that it is on average, lower for informative paths.
Moreover, $V_P$ proves to be a reliable measure of the noise of the paths. Indeed, it is, on average, three orders lower in the paths from Xu et al.'s synthetic data in comparison with the synthetic data reproducing the complexity of connections and dynamics of real datasets, like the Orbis and flights datasets, thus being substantially more noisy (see Table\ref{tab:noise}). Also, as expected, it has a an inverse relation with the $F_{P_H}$ and $F_{P_L}$ (see Fig.s \ref{fig:D_F},\ref{fig:D_F_Flight}). Indeed, the statistics and dynamics of paths appearing a higher amount of times in the data are more defined and, consequently, less noisy. Further details on the measures analysis can be found in Supplementary Information, Section \ref{subsec:distances_analysis}.

\subsection{Real-world Data: Ownership Sequences from the Orbis Dataset}

After showing the performance of our model in synthetic data, we showcase the results of DIVOP and HYPA when applied to a real-world dataset: international ownership sequences of multinational corporations. Ownership sequences track the ownership structure from subsidiaries in specific countries to the ultimate parent entity. The states represent the countries where the subsidiaries and owners are located, and the edges represent majority ownership stakes. These sequences reflect financial flows in the form of dividens and reflect how multinational corporations organize their structure to take advantage of beneficial legislation (for example, to minimize tax payments \cite{garcia2017}).

In this section, we focus on the role of tax havens in ownership sequences. We use the definition of tax havens by \cite{garcia2017}. Of the 364,811 ownership sequences analyzed, 46\% contain at least one tax haven. Of all paths (sub-sequences) of order 2 to 5, 68\% contain at least one tax haven.

\textbf{Informative paths:} Our model detected 472 higher-order nodes, in contrast to 1,132 identified by the HYPA model. Of the 472 higher-order nodes, 386 (82\%) included at least one tax haven. In the HYPA case, 72\%  of the higher-order nodes included at least one tax haven. In contrast to the first-order dynamics, in informative paths every step is influenced by the previous one. This means that the presence of some countries appearing before the last state of the informative path manages to change its strategy in locating its subsidiaries in different legislations, compared to the case in which the path is not informative. Consequently, having a higher percentage of tax havens in informative paths signals that the presence of tax havens in a path is linked to a higher chance to influence the dynamics of firms' locations in a different direction, changing the sites of the subsidiaries owned by the firm at the end of the path. This finding indicates a strategic role played by tax havens in the construction of corporate structures.

A key difference between DIVOP and HYPA is that while HYPA finds mostly second-order informative paths, DIVOP detects mostly third and fourth order paths. In particular:
\begin{itemize}
    \item All the paths (informative and not) are 14\% of the second order, 30\% of the third, 31\% of the fourth, and 25\% of the fifth order.
    \item DIVOP detected 21\% dependencies of the second order, 35\% of the third, 28\% of the fourth, and 16\% of the fifth order.
    \item The HYPA model detected a higher proportion of second-order informative paths (50\%), 36\% of the third, and 14\% of the fourth order.
\end{itemize}

This distribution highlights HYPA's focus on lower order dependencies (a tendency revealed in synthetic data as well) and suggests our model might better capture the higher-order interactions typically overlooked, given that there is a higher correspondence between the percentage of the orders of paths present in the data and the order of the ones detected by our model.

The potential bias of HYPA towards second-order paths can also explain the difference in the number of paths detected by DIVOP and HYPA. Only 167 paths (35\% of the paths found by DIVOP) match the ones found by HYPA, while this number was 75\% in synthetic data. However, in synthetic data the majority of the informative paths were second-order paths, where the performance of both algorithms is comparable (Fig.\ref{fig:PR_O}). In real Orbis data, 540 of the 1,132 paths found by HYPA are segments of the paths detected by DIVOP. This indicates that while HYPA struggles with higher-order dependencies, it still recognizes elements that are part of more complex structures. Indeed, the ``anomaly'' of the higher-order paths' distribution probably propagates to the distribution of the corresponding lower order and this is what is detected by HYPA. For instance, if $(A,B,C)$ is a chain but $(B,C)$ is not, $D(B,C)$ is affected in the anomaly by $D(A,B,C)$, in particular by the $(B,C)$ repetitions by the $A$ state.
Interestingly, of the matching paths, 85\% contain at least one tax haven, which is a higher percentage compared to the ones contained in DIVOP or HYPA alone.

\textbf{Mixed-order network structure:} After detecting the informative paths using DIVOP and HYPA, we can reconstruct a mixed-order network following the representation method in \cite{Xu2016}. In the mixed-order netowrk, nodes represent ownership paths of multinational corporations. This mixed-order network  represents the real-world sequences more accurately, improving the analysis of centralities and clustering \cite{Xu2016,Rosvall2014,salnikov2016using}. For comparison, we also construct a first-order network.

We start studying the most important metrics. For both HYPA and DIVOP networks, we find that China plays a very important role, being ranked 1\textsuperscript{st}--3\textsuperscript{rd} in PageRank (PR), Betweeness Centrality (BC) and Eigenvector Centrality (EC) in both networks. This result is in contrast to the one found in the first order network, in which China covers either the 7\textsuperscript{th} place (EC) or the 16\textsuperscript{th} (PR and BC). Among the informative paths, the ones containing European tax havens (Great Britain, The Netherlands, Luxembourg, etc...) are the most important in the aforementioned centralities. These results confirm the major role of tax havens in the corporate structures of multinationals. 
A major role as first-order nodes, instead, is covered by the United States (as expected) and, interestingly, also by ``YY'', which stands for unknown/unreported country in all the three networks. Clustering through modularity, the node YY is consistently placed in the community containing both Asian countries and small island tax havens in the three networks, hinting at its role in taxation and possibly information concealment \cite{schjelderup2016secrecy,christensen2012hidden}. 

\textbf{Random walk through the reconstructed networks:} To further understand if DIVOP and HYPA reconstruct reliable dynamics through the mixed-order network, we test them through a random walk, following the reconstructed probability distributions among the connected nodes. The test consists in letting an agent walk through the sequences dataset and choose the next step with the probability reported in the tested network. If an informative path appears in the data as part of a sequence, the random walker will follow the higher-order distribution for the next step instead of the first order one. We have tested the ability of the models to forecast an unknown part of the dataset, dividing it into training and test sets. The k-cross validation will be with $k=5$ and we use the training set (4 of the 5 parts of the data) to create the networks and then let the agent walk in the test set (the remaining fifth). We perform 5 repetitions, every time with a different fifth of the dataset as the test set. DIVOP and HYPA show a similar result, as can be seen in Table\ref{tab:random_walk}, but considering that DIVOP obtains the same percentage with a lower number of nodes. The performance of the random walker is significantly improved compared to the first order network.

\begin{table}[htbp]
\centering
\begin{tabular}{lccc}
\toprule
                & DIVOP               & HYPA               & First Order        \\
\midrule
\textbf{Forecasting RW} & $0.249 \pm 0.001$ & $0.249 \pm 0.002$ & $0.197 \pm 0.002$ \\
\bottomrule
\end{tabular}

\vspace{0.2cm}

\caption{\textbf{Performance of the random walker on the real Orbis dataset.} 
 Correct rate of steps performed by the random walker following the transition probabilities of the networks built through DIVOP, HYPA, and first-order models. Dynamics reconstructed on 4/5 of the dataset and tested on the remaining fifth, repeated 5 times.}
\label{tab:random_walk}
\end{table}

\begin{figure}[t]
    \centering
    \includegraphics[width=0.5\textwidth]{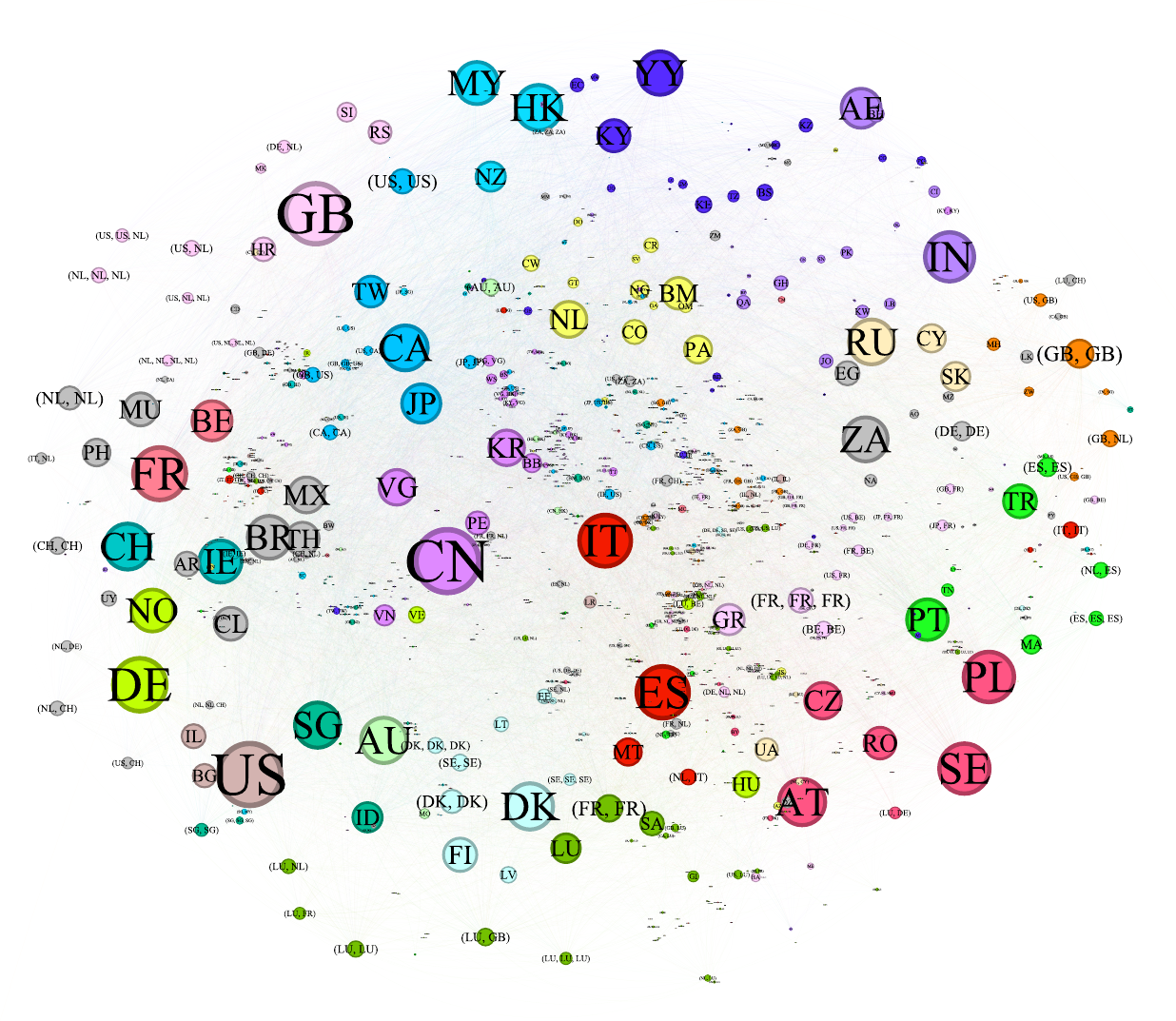}
    \caption{\textbf{Orbis real dataset DIVOP network:} Representation of the reconstructed variable-order network through Xu et al. \cite{Xu2016} method. Nodes are the locations of firms in countries labeled with ISO-2 code. Higher-order nodes are informative paths and the ownership of firms goes from the left to the right, e.g. node (US,NL) represents the couples of firms in which the owner is located in the US, while its subsidiary is in The Netherlands. The edges represent ownership relations between nodes. Communities are obtained through modularity and highlighted through different colors. The used layout is ForceAtlas, Gephi.}
    \label{fig:O_OM_raw}
\end{figure}



\section{Conclusion}\label{sec:conclusion}

In this paper, we address the challenge of overfitting in variable-order Markov models based on the Kullback-Leibler (KL) divergence. Overfitting includes fluctuations in the dynamics reconstruction, which hinders the most efficient
forecasting of transitions and the allocation of resources in the most
efficient way. To reduce overfitting, we introduce a measure of noise through cross-validation into traditional methods based on the KL divergence to detect higher-order informative paths in sequences data and construct variable-order Markov models.
The key innovation lies in measuring the variability of the distribution of destinations ($V_{P_H}$  and $V_{P_L}$) between the training and validation sets, which allows the algorithm to model how likely the divergence between the lower- and higher-order distributions is to be driven by noise. This measure proves to be crucial in distinguishing the higher-order dependencies and a reliable method to measure the noise of paths inside datasets.
We systematically evaluate the performance of our algorithm, DIVOP, in a variety of synthetic datasets, showing high precision and recall in all cases. Additionally, we test the ability to forecast the dynamics on a real dataset of corporate structures, through a random walk test. In both cases we obtain better results in comparison with other higher- and first-order models. 
Moreover, from the analysis of the real Orbis dataset, we find how tax havens play a key role in the organization of multinational corporations' structures. Indeed, the presence of tax havens in paths is linked to a higher chance to change the dynamics of the structures. Furthermore, the most important paths at the level of centralities in the network are those containing European tax havens.

The second innovation of this paper is the creation of synthetic data that resembles real-world datasets. This synthetic data construction allows us to compare the performance of different models and to adapt our and other models' thresholds/algorithms to the dataset characteristics. Our open-source code facilitates reproducibility and invites collaboration.

Our model can be applied to a variety of real-world sequence data, providing better representations of the structure and dynamics in the data. This is, for example, useful to detect and understand the structural properties of a network, such as the clusters of states present in the data. When sequence data represent time series (e.g., in the flight dataset), mixed-order models allow us to better represent the dynamics. For example, if we find that passengers from the United States flying to the United Kingdom will more likely fly afterward to Italy (i.e., there are higher order dependencies), the mixed-order model would indicate that it would be beneficial to organize the timing of flight transfers accordingly or to increase the frequency of flights from the United States to Italy. By avoiding overfitting, our model prevents the creation of "false" and unnecessary transfers (noise mistaken for dependencies in flights) and the waste of resources.

Our algorithm could also be applied to single sequences  (e.g. an individual DNA sequence), allowing the identification of subsequences (paths) with specific functions---for example, protein binding sites.

We see one avenue of future work as especially promising. Our results show that any algorithm using a simple threshold to distinguish informative paths will be suboptimal, since the threshold depends on the frequency of the paths, number of nodes in the first-order network, and size of the dataset. We circumvent this limitation by fitting our classifier to synthetic data including a variety of path anomalies and data sizes. Future work could create a general classifier based on a larger variety of synthetic data. 

\backmatter

\bmhead{Acknowledgements}

The authors acknowledge the support from Czech Science Foundation (CORPTAX, 21-05547M). Valeria Secchini acknowledges support from the Horizon Europe project `DemoTrans' (grant 101059288), support from the Charles University Grant Agency (GAUK, 254322) and support from the Centre for Complex Systems Studies of Utrecht University (Swaantje Mondt Fund).




\newpage

\begin{appendices}

\section{Methodology and Data: Additional Information}

\subsection{Data}

We base our analysis on three datasets.

\begin{itemize}

\item[-]\textbf{\textit{Orbis data on cross-national corporate ownership sequences:}}
Orbis is a widely used source that encompasses over 300 million public and private companies globally. We extract corporate ownership sequences of multinational corporations worldwide. For each subsidiary of multinational corporations, we construct the sequence of ownership from the subsidiary to the top holding, keeping only majority owned relationships (over 50\% ownership). We use data extracted from 2017 for $18,825$ multinationals, and create all corporate ownership sequences ($364,811$ sequences) representing cross-border ownership relationships between $206$ states (countries). The distribution of the length of the sequences is skewed, with the majority of the sequences having length $2$ and the average length being $3.52$. The density of the adjacency matrix of the first order network is $0.15$. We analyze this data with all algorithms, and use its characteristics to create synthetic data (Section \ref{sec:methods:syn})

\item[-]\textbf{\textit{Flying Data:}}
This dataset contains information on flight routes through 354 US airports, collected by \cite{Flights_data}. It contains $185'871$ sequences, $382$ nodes, with sequences having an average length of $4.02$ and a median of $5$. The adjacency matrix has a density of $\sim 0.05$. For more information, see \cite{Larock2020}. Similarly to the Orbis case, we create synthetic data resembling the Flying dataset.

\item[-]\textbf{\textit{Xu et al. synthetic data:}} 
This dataset is used in \cite{Xu2016} to test their algorithm. There are $100$ possible nodes in $100,000$ sequences, each with a length of $100$ steps. The dataset contains $30$ synthetic higher-order nodes.

\end{itemize}

\subsection{Related models to extract informative paths (higher-order nodes in variable-order MCM)}\label{Other_models}
We compare DIVOP to two recently developed algorithms to extract informative paths from sequences data.
\begin{itemize}
    
\item[-]\textbf{{\textit{BuildHON+ and BuildHON: Xu/Saebi et al. \cite{Saebi2020,Xu2016}}}}
DIVOP is closely based on BuildHON/BuildHON+. Both BuildHON and BuilHON+ compare the probability distribution of the path of order $k$ ($P_H$) with its lower order counterpart $k-1$ path ($P_L$) distribution using the KL divergence. Starting from second-order paths, BuilHON includes the second-order paths if the KL divergence is higher than a certain threshold, $\delta$. After this step, it considers paths of third order and compares them with the corresponding second order paths if they have not been pruned, otherwise with their corresponding first order. BuildHON+ proceedes with this process only if the maximum possible KL divergence for each $k-1$ order path is higher than $\delta$ when increasing the order. In this way it decreases the amount of analyzed paths and it stops automatically. 
The threshold is a parameter of the model, and the suggested default is $\delta = \frac{k}{log_2(F_{P_k})}$ (BuildHON) and $\delta = \frac{k}{log_2(1+F_{P_{k}})}$ (BuildHON+). In both cases, the threshold is proportional to the order and inversely dependent on the frequency with which the path appears. This is done to increase the threshold for paths that occur rarely and to reduce their overfitting.

\item[-]\textbf{{\textit{HYPA: LaRock et al. \cite{Larock2020} }}}
HYPA model finds \textit{anomalous} paths (which will later become the higher order nodes of the Markov chain) through a null model. They construct De Bruijn graphs for each order $k$, where each graph contains every possible $k$th-order node and they are connected to each other considering their transition probabilities. They take into account two variables: 
\begin{itemize}
    \item[$\boldsymbol{\cdot}$] $f$, being the real weight of the edge between two paths of order $k$ in the $k$-th De Bruijn graph (the starting path would be the one pointing at the other path through the edge).
    \item[$\boldsymbol{\cdot}$] $X$, being the random variable assuming the weight of the edge between the two paths in a random realization of a $k$-th order De Bruin graph, starting from the graph of order $k-1$.
\end{itemize}

They estimate the HYPA$^{k}$ score of the aforementioned edge comparing these two quantities. In particular, it is the probability that $X$ is lower or equal to $f$. If this probability is near to $0$ or to $1$, meaning the expected weight is distant from the real one, then the ``starting'' path is considered anomalous. The threshold of the HYPA$^{k}$ score is not defined. In our analysis, we set it through our synthetic data, such that it maximizes the F1 score in finding the synthetic higher order nodes.

\end{itemize}

\section{Results: Additional Information and Plots}

\subsection{Impact of the metrics in the classification algorithm}\label{sec:variable_analysis}
\begin{figure}[t]
    \centering
    \includegraphics[width=0.5\textwidth]{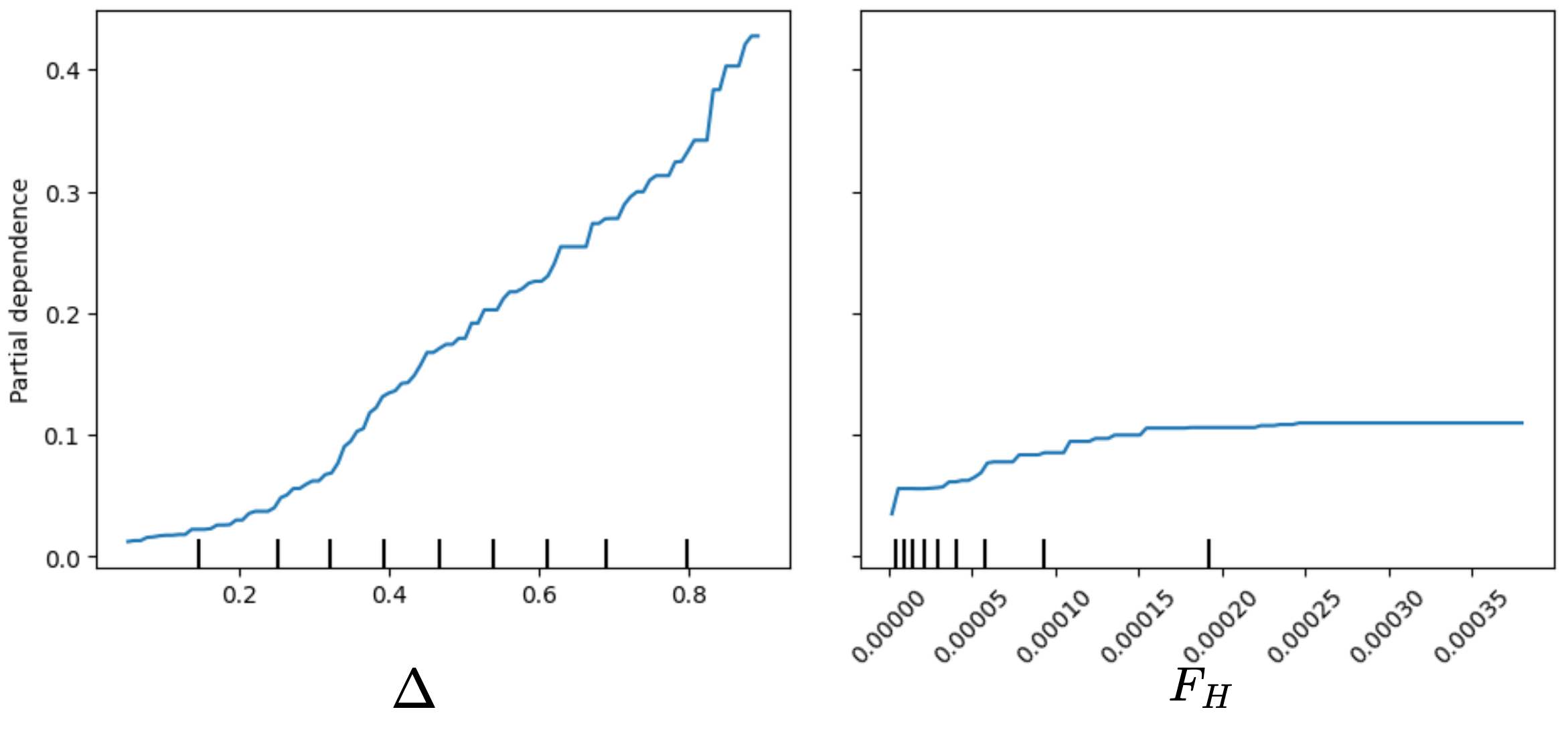}
    \caption{Partial dependence of (left) $\overline{\Delta}$ (right) $F_{P_H}$}
    \label{fig:HMFh}
\end{figure}

\begin{figure}[t]
    \centering
    \includegraphics[width=0.5\textwidth]{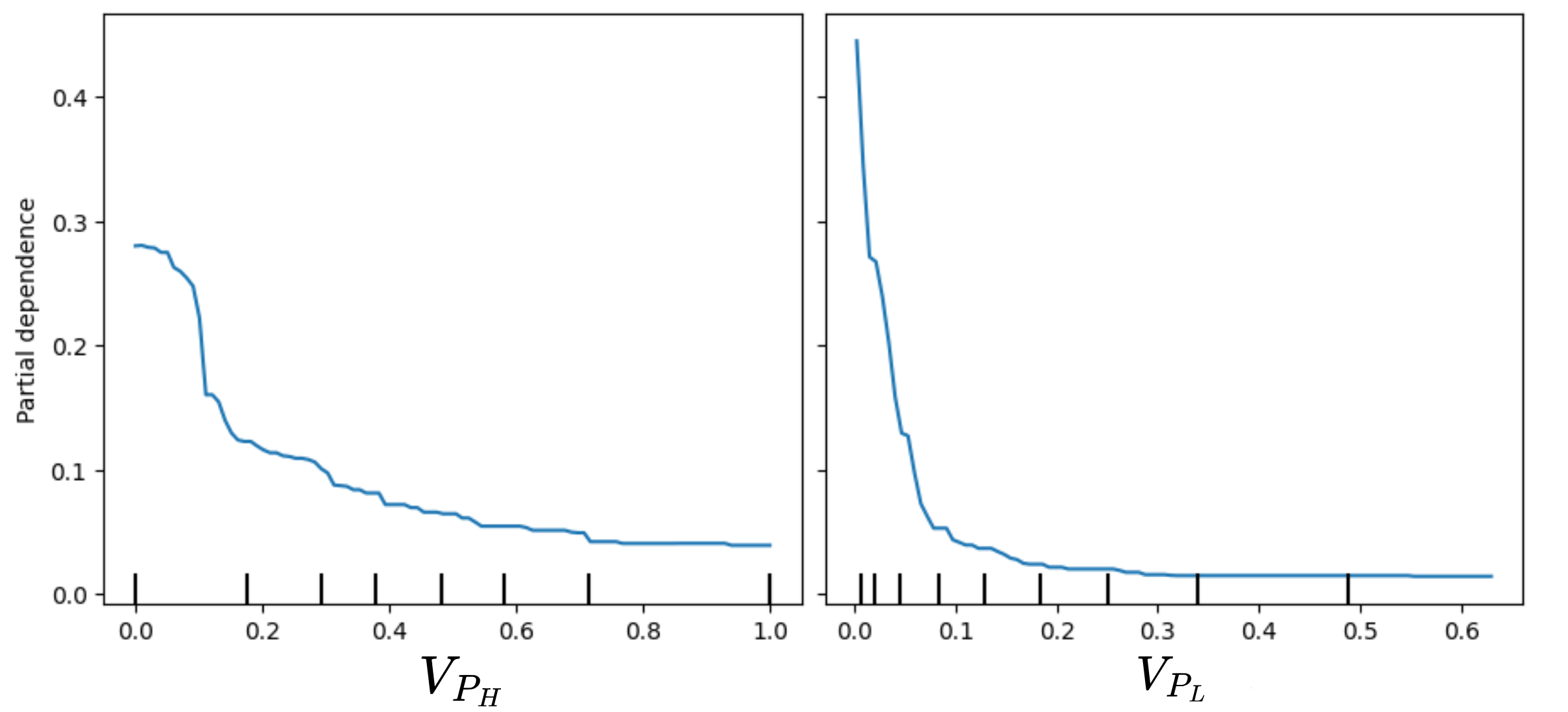}
    \caption{Partial dependence of (left) $\overline{V_{P_H}}$ (right) $\overline{V_{P_L}}$}
    \label{fig:SIVs}
\end{figure}

Here we will show, in the case of synthetic data inspired by Orbis data, how changes in the metrics affect the classification of paths as informative. We do this using partial dependence plots \cite{molnar2020interpretable}. These plots show a non linear relationship between the predicted outcome and the analyzed quantities. More specifically, the increase in the $\Delta$ (\ref{fig:HMFh}A) shows a higher partial dependence on the predicted outcome and the same happens for $F_H$(\ref{fig:HMFh}B). The opposite is shown for the paths variability (\ref{fig:SIVs}).

\subsection{Disaggregating the performance of the algorithms} \label{subsec:disaggregating}
After showing the global performance  of all models, we then analyzed the precision and recall of each model disaggregating the results for both the Orbis and Flights dataset by the path \textit{order} and \textit{frequency} of the path.

\subsubsection*{By path frequency}
We first analyzed the performance of all algorithms depending on the frequency of the paths, irrespectively of path order. We divided the paths according to the path frequency in nine groups of equal number of cases. In terms of precision (Fig.~\ref{fig:D_F}A), our algorithm performs comparable or better than HYPA for all cases, and especially so in the low frequency cases where the path was observed less than 20 times. The precision of DIVOP and HYPA ranged from $\sim 95\%$ (for both models) to $\sim 50\%$ for HYPA and $\sim 70\%$ for DIVOP for paths observed 4 or 5 times, to $\approx 90\%$ for paths observed at least 50 times. The buildHON algorithm performed poorly for all path frequency categories, indicating a high number of false positives. Interestingly, the precision of buildHON is higher for paths observed 5 or 6 times. This is due to how the classifier is set in that model. As explained in Section\ref{Other_models}, the threshold decreases as the frequency (called \textit{Support} in \cite{Saebi2020}) increases. This attempts to account for the fact that the noise is lower at higher frequencies. However, this adjustment creates a problem of overfitting at higher frequencies. 
The recall of all algorithms is similarly reduced as the frequency decreases (Fig.~\ref{fig:D_F}B). While the recall is $\approx 80\%$ for paths observed more than 50 times, it decreases to $\approx30\%$ for paths observed 4 or 5 times. 

\begin{figure}[h]
    \centering
    \includegraphics[width=0.5\textwidth]{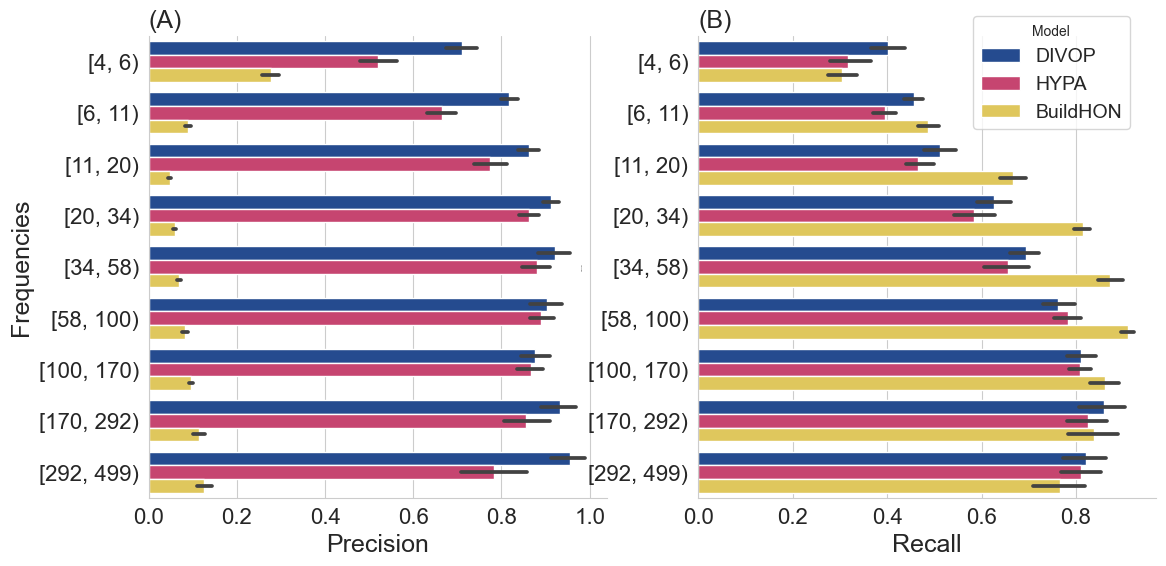}
    \caption{\textbf{Algorithm performance by path frequency} on the Orbis dataset. For each dataset, the performance of DIVOP (blue), HYPA (red) and buildHON (yellow) in terms of precision (A) and recall (B) is shown. The analysis on the Flights data, showing comparable results, can be found in Fig. \ref{fig:PR_F_Flight}.}
    \label{fig:PR_F}
\end{figure}

\begin{figure}[h!]
    \centering
    \includegraphics[width=0.5\textwidth]{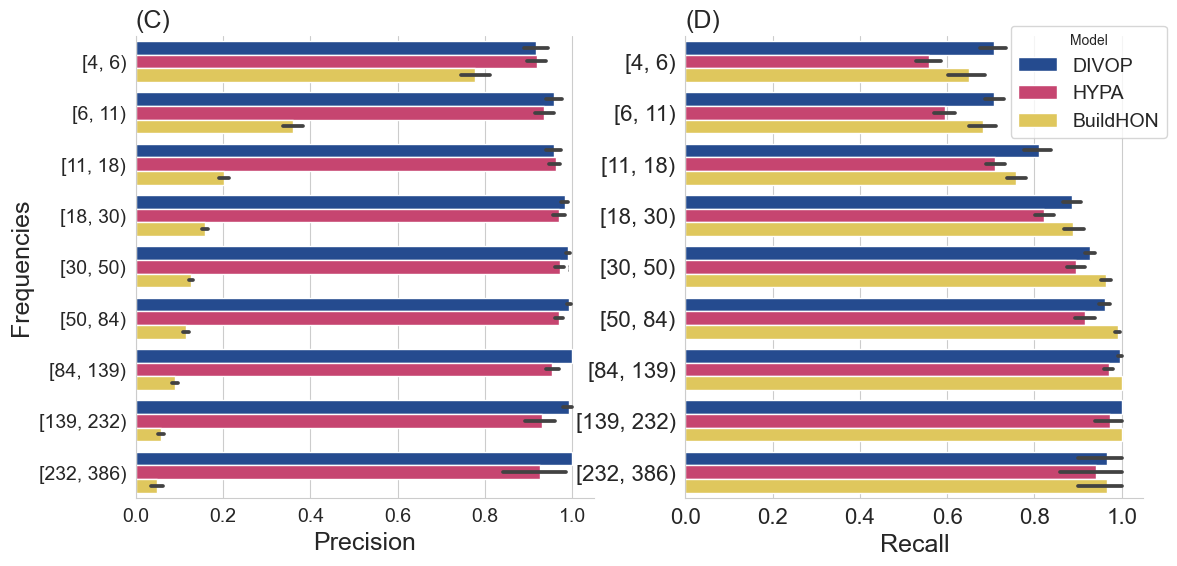}
    \caption{\textbf{Algorithm performance by path frequency} on the flight dataset. Precision and recall for different numbers of appearances in the data for every model, flights data. Recall decreases with the decreasing of the frequency, Precision slightly changes. In this case as well, the precision of the Saebi model decreases with the increasing of the frequency because of how the threshold for the KL divergence is built in the model.}
    \label{fig:PR_F_Flight}
\end{figure}

\subsubsection*{By path order}
\begin{figure}[h!]
    \centering
    \begin{minipage}{0.5\textwidth}
    \includegraphics[width=1.05\textwidth]{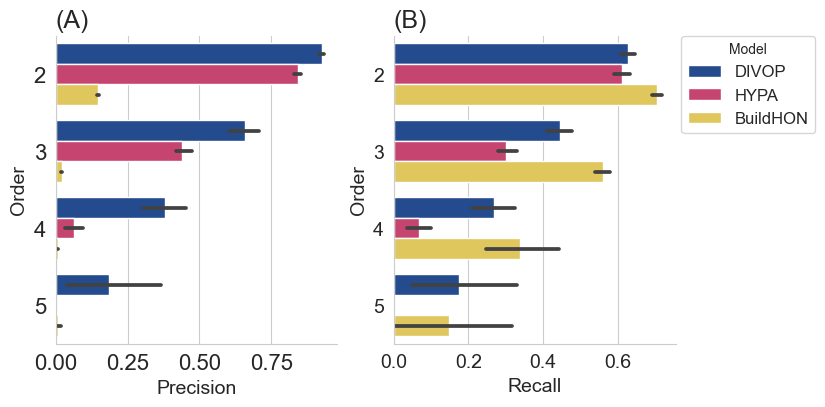}
    \vspace{-0.5cm}
    \caption*{Figure 8.1: (a) Orbis Data}
    \end{minipage}\hfill
    \begin{minipage}{0.5\textwidth}
    \includegraphics[width=0.85\textwidth]{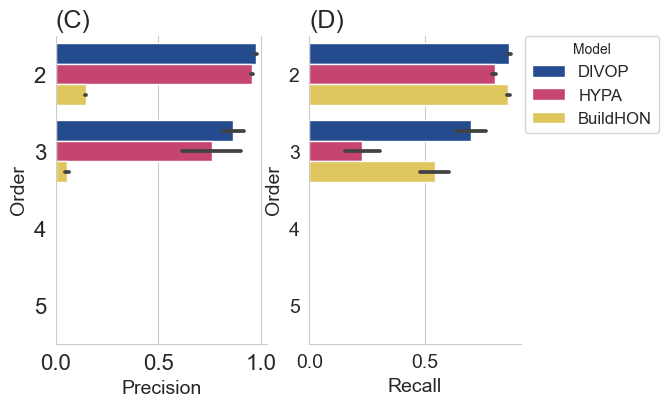}
    \caption*{Figure 8.2: (b) Flights data}
    \end{minipage}\hfill
    \caption{\textbf{Algorithm performance by path order} on the (A--B) Orbis (C--D) Flights datasets. For each dataset, the performance of DIVOP (blue), HYPA (red) and buildHON (yellow) in terms of (A,C) precision and recall (B,D) is shown. Note that fifth order nodes are not present in the HYPA metrics, due to computational limitations.}
    \label{fig:PR_O}
\end{figure}

For second-order nodes-paths(e.g. UK|IT|NL), DIVOP and HYPA performed similarly in both precision (Fig.~\ref{fig:PR_O}A,C) and recall (Fig.~\ref{fig:PR_O}B,D), with measuring ranging from 60 to 95\%. 
For paths of order three and above, both the precision and recall decrease, especially in HYPA. This is expected, as the frequency of those paths is lower, and the algorithms' performance declined in those cases. 
For all orders, the buildHON algorithm performed slightly better in terms of recall (Fig.~\ref{fig:PR_O}B,D), but much worse in terms of precision (Fig.~\ref{fig:PR_O}A,C). 

\subsection{The measure of noise: analysis and meaning}\label{subsec:distances_analysis}

\begin{figure}[h!]
    \centering
    \begin{minipage}{0.5\textwidth}
    \includegraphics[width=0.9\textwidth]{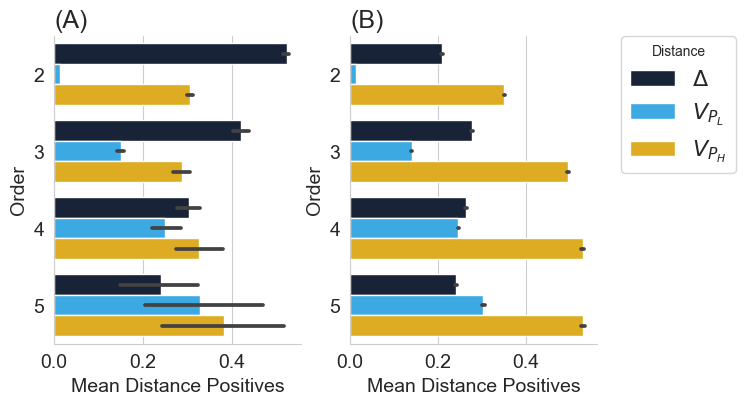}
    \caption*{Figure 9.1: Orbis Data}
    \end{minipage}\hfill
    \begin{minipage}{0.5\textwidth}
    \includegraphics[width=0.9\textwidth]{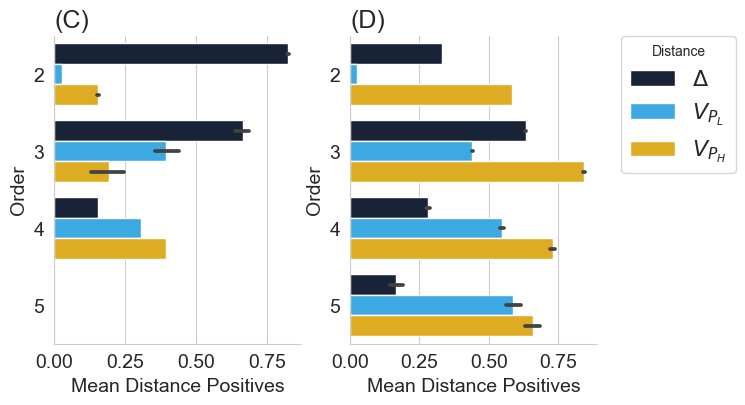}
    \caption*{Figure 9.2: Flights data}
    \end{minipage}\hfill
    \caption{$\Delta$, $V_{P_L}$ and $V_{P_H}$ at different orders both for flights data (A-B) and Orbis data (C-D), both for positive and negative paths. $\Delta$ is significantly different between negatives and positives mostly for the second order, which makes the two $V_P$s quite important to detect the informative paths. The noise increases with the order, as we can see from the $V_P$s, given that there is a lower number of appearances at higher-orders. The error in the fourth order for positives in (C) cannot be calculated given that we have one synthetic informative path only for that order. We do not have statistics for the fifth order given that there were no fifth order synthetic higher-order dependencies created in the data for (C).}
    \label{fig:D_O}
\end{figure}

\begin{figure}[h!]
    \centering
    \includegraphics[width=0.5\textwidth]{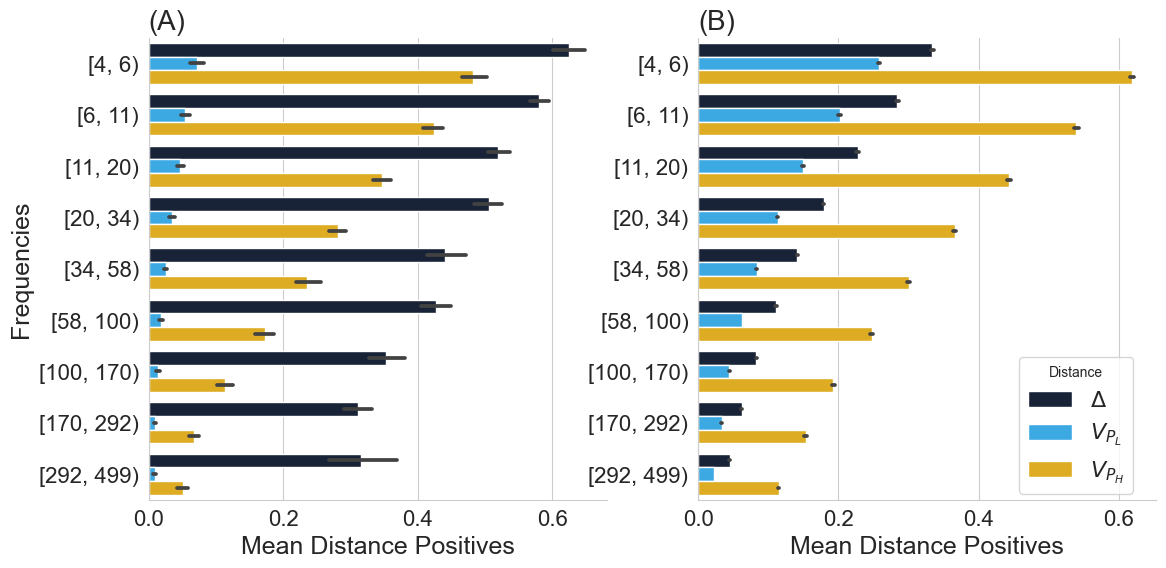}
    \caption{$\Delta$, $V_{P_L}$ and $V_{P_H}$ at different frequencies for Orbis data both for positive and negative paths. The noise decreases with a higher number of appearances of the paths in the data, as we can see from both $V_{P_H}$ and $V_{P_L}$, and as we would expect. $\Delta$ is higher for positives than for negatives, as expected. The general decreasing of $\Delta$ with the increasing of the frequency, is because a higher noise creates a higher variability in the distributions.}
    \label{fig:D_F}
\end{figure}

\begin{figure}[H]
    \centering
    \includegraphics[width=0.5\textwidth]{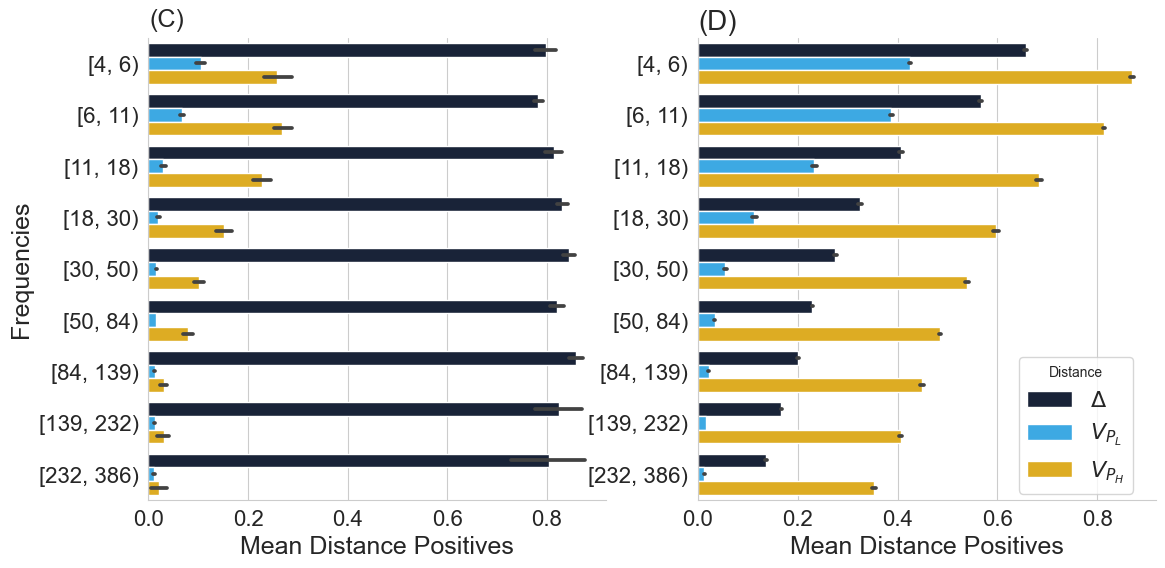}
    \caption{$\Delta$, $V_{P_H}$ and $V_{P_L}$ at different frequencies for flights data, both for positive and negative patterns. The noise decreases with a higher number of appearances of the patterns in the data, as we can see from both $V_{P_H}$ and $V_{P_L}$, and as we would expect. $\Delta$ is higher for positives than for negatives, as expected. $\Delta$ decreases with the increasing of the frequency for negatives, but not for positives, probably because that the positives' noise does not change significantly with the frequency.}
    \label{fig:D_F_Flight}
\end{figure}

In Fig.s \ref{fig:D_O} and \ref{fig:D_F} we show the behavior of the $V_{P_L}$, $V_{P_H}$ and $\Delta$ both for synthetic paths that are informative (the positives) and for the ones that are not (the negatives). The quantities are shown at different orders in Fig. \ref{fig:D_O} and at different frequencies in Fig. \ref{fig:D_F} and Fig. \ref{fig:D_F_Flight}, for both Orbis synthetic data and for Flight synthetic data. $\Delta$ is significantly different between negatives and positives mostly for the second order, while it looks more comparable at higher-orders, which makes the two $V_P$s rather important to detect the informative paths at higher-orders. The noise increases with the order, as we can see from the $V_P$s, given that there is a lower number of appearances of the path at higher-orders. As regards the frequencies, the noise decreases with an higher number of appearances of the paths in the data, as we can see from both $V_{P_H}$ and $V_{P_L}$, and as we would expect. $\Delta$ is higher for positives than for negatives, as expected (informative paths bring more information than negative paths). The general decreasing of $\Delta$ with the increasing of the frequency, is because that the higher amount of noise creates a higher variability in the distributions, which also makes the lower and higher-order distributions distant among them. This reinforces the idea that $V_P$s are crucial to distinguish real additional information from noise. Interestingly, for the flights data there is not a decrease of $\Delta$ with the increase of the frequency for positives. This could be because there is a low amount of noise in the informative paths and it does not decrease significantly with frequencies.

Looking at Table \ref{tab:noise}, a general trend shows that the $V_P$s are significantly higher for negatives in all the three datasets analyzed. The difference in the $V_P$s between the positive and negatives is because informative paths have, in general, a less uniform distribution and (more likely) with a lower number of destinations, implying a lower variability between the two distributions in the validation and training set. This "peaking" in the distributions when including higher-orders in the network (which means less variability in the probability of the destinations) can also be seen in \cite{Rosvall2014}, where they find that the entropy of the network decreases in second order dynamics in comparison with the first one, because the uncertainty of the next step of the flow is lower in a random walk, given the current state. If the second order paths were not more informative compared to the first one, i.e. the second order can also be described by the first one, there would be no difference in the entropy. So the informative paths of the second order are the cause of the entropy decrease and not the general second order paths. The negatives have a distribution that is more uniform/with higher uncertainty in comparison with the positives, because it is more similar to the distribution of its lower order. As expected, in the synthetic data from Xu et al. \cite{Xu2016}, the noise is substantially lower when compared  to the one in Orbis and flight synthetic data. This is because in Xu dataset the connections have fewer destinations (only 4 for each first order state, even fewer for higher order ones), being created in a standard way for each state. This procedure implies a very low amount of fluctuations in comparison with our method which reproduces the complexity of connections and dynamics of real datasets. 

  \begin{table}[htbp]

    	\centering
    	\begin{tabular}{lllll}
    		\toprule
    		
    	   Dataset   & \multicolumn{2}{c}{Average $V_{P_H}$}  & \multicolumn{2}{c}{Average $V_{P_L}$}\\
    		\midrule
      &Positives&Negatives&Positives&Negatives\\
                \midrule

    Orbis & $0.305 \pm 0.009$   & $0.494 \pm 0.004$  & $0.040 \pm 0.003$  & $0.185 \pm 0.002$  \\
    Flights & $0.156 \pm 0.008$   & $0.748 \pm 0.003$  & $0.037 \pm 0.003$ & $0.303 \pm 0.001$  \\
    Xu  & $(2.3 \pm 3)\cdot 10^{-5}$   & $0.026 \pm 0.002$  & $(1.9 \pm 3)\cdot 10^{-5}$ &  $0.006 \pm 0.001$ \\

    		\bottomrule
    	\end{tabular}
     
         	\caption{\textbf{Noise average measures for higher- ($V_{P_H}$) and lower- ($V_{P_L}$) order paths}. We show the average variability of paths in the three different datasets for both positives (informative paths) and negatives. The paths are divided into lower and higher order.}
    	\label{tab:noise}
    \end{table}


In conclusion, $V_P$ proves to be a good measure of noise, also considering how it is generally significantly lower in the Xu et al. synthetic data, in comparison with the synthetic data reproducing the complexity of connections and dynamics of real datasets, like the Orbis and flights datasets. Further, as expected, it has a an inverse relation with the frequencies of appearances of the paths. Indeed, the statistics and dynamics of paths appearing a high amount of times in the data are more defined and, consequently, less noisy. We have also shown how the variability is crucial in pruning models to distinguish the informative paths, given that, for orders higher than two, the average $\Delta$ is quite similar between positives and negatives. The variability, by contrast, is on average lower for informative paths.

\vspace{1.5cm}

\end{appendices}


\bibliography{sn-bibliography}

\end{document}